\documentclass[11pt,a4paper]{article}
\usepackage{epsfig}

\title{ Quantum Fluctuations of a Single Trapped Atom:
        Transient Rabi Oscillations and Magnetic Bistability}

\author{H. Schadwinkel, V. Gomer, U. Reiter, B. Ueberholz, and D. Meschede\\
        Institut f\"ur Angewandte Physik, Universit\"at Bonn,\\
        Wegelerstr. 8, D-53115 Bonn, Germany
        }

\begin{document}
\maketitle
\begin{abstract}

 Isolation of a single atomic particle and monitoring its resonance fluorescence is a
 powerful tool for studies of quantum effects in radiation-matter interaction. Here we
 present observations of quantum dynamics of an isolated neutral atom stored in a
 magneto-optical trap. By means of photon correlations in the atom's resonance
 fluorescence we demonstrate the well-known phenomenon of photon antibunching which
 corresponds to transient Rabi oscillations in the atom. Through polarization-sensitive
 photon correlations we show a novel example of resolved quantum fluctuations: spontaneous
 magnetic orientation of an atom. These effects can only be observed with a single atom.

\end{abstract}

keywords: laser cooling and trapping, single atoms, photon
correlations, quantum fluctuations

\section{Introduction}

 Radiation matter interaction has been studied for a long time with atomic samples. It is
 known that density matrix theory is well suited to exhaustively describe the properties
 of fluorescence from macroscopic atomic ensembles, and that in this limit semiclassical
 theory gives excellent approximations, since phenomena related to the quantum nature
 of the light field are often hidden.

 On the other hand, it has also been realized since more than two decades now that
 isolation of a single atom provides an opportunity to directly observe pure quantum
 properties of the interacting radiation-matter system. A celebrated example is the
 observation of photon antibunching, where initial experiments \cite{Kimble77,Rateike82}
 were carried out with extremely diluted atomic beams, and more precise investigations
 became possible when ion traps were employed to achieve long-term confinement of a single
 atomic particle \cite{Diedrich87}.

 For neutral atoms the confinement strength is reduced since trapping has to rely on
 forces derived from electric or magnetic dipole interaction. Therefore neutral atoms have
 only more recently become available for measurements at the microscopic level through the
 application of laser cooling techniques well established for macroscopic samples with
 many atoms. With laser light single neutral atoms can be trapped and observed for long
 times \cite{Gomer98a,Gomer98b} in analogy with ion traps.

 One of the most powerful methods to observe quantum fluctuations in atom-radiation
 interaction is to measure photon correlations of the light emitted by a single atom. The
 quantum theoretical description of photodetection was put forward by Glauber
 \cite{Glauber63} in the early 60s, and very successfully applied to all appropriate
 experiments since. In the photon language the normalized intensity-intensity correlation
 function is the conditional probability to detect a second photon if a first one was
 detected a time $\tau$ before. This second order correlation function is defined by

\begin{equation}\label{eq:g2}
        g^{(2)}(\tau) =
        \frac{\langle : \hat{n} (t+\tau)\hat{n}(t) : \rangle }
        {\langle \hat{n} (t) \rangle\langle \hat{n}  (t)
        \rangle}\quad ,
\end{equation}
\noindent
 where $\hat{n}(t)=\hat{a}^\dagger(t)\hat{a}(t)$ is the photon number operator constructed
 from field operators $\hat{a}^\dagger$, $\hat{a}$, and where $: \hat{n}(t+\tau)\hat{n}(t) :$
 denotes normal ordering of field operators. At sufficiently long delays $\tau$ all possible
 correlations have decayed, and hence
 $\langle :\hat{n}(t+\tau)\hat{n} (t) :\rangle \to \langle \hat{n} (t) \rangle \langle \hat{n} (t) \rangle$.
 It is thus often informative to discuss the deviation of $g^{(2)}(\tau)$ from unity,
 or the quantity $\mid g^{(2)}(\tau)\!-\!1\mid$ as a measure of the fluctuation strength
 of the system, called \emph{contrast} in the following.

 For auto correlations (\ref{eq:g2}) all classical fields must obey $g^{(2)}(0)\geq 1$,
 while photon antibunching shows $g^{(2)}(0)=0$, making it a prime example of quantum
 field fluctuations. In an intuitive interpretation it is said \cite{Walls94} that the
 first detected photon projects an atom in its ground state. From this initial state the
 atom then relaxes back to its equilibrium state on the time scale of its radiative decay.
 In this sense the observation of this fluctuation is 'measurement induced'.

 In our work we are additionally interested in the case where different atomic transitions
 can be distinguished by their polarization properties. We therefore analyze cross
 correlations of orthogonal polarization states $\mid\beta\rangle$ and $\mid\alpha\rangle$
 of the resonance fluorescence from a single atom,

\begin{equation}\label{eq1:g2}
        g^{(2)}_{\alpha\beta}(\tau) =
        \frac{\langle : \hat{n}_\alpha (t+\tau)\hat{n}_\beta(t) : \rangle }
        {\langle \hat{n}_\alpha (t) \rangle\langle \hat{n}_\beta (t)
        \rangle}\quad .
\end{equation}
\noindent
 A single trapped Cesium atom shows very strong
 polarization correlations in its resonance fluorescence.
 While this effect is very pronounced for correlations
 between orthogonal circular polarizations of the detected
 photons, it vanishes for linearly polarized photons even in
 a light field with linear polarization at every place. It
 is the purpose of this work to show that the strong
 contrast in $g^{(2)}_{\alpha\beta}(\tau)$ that we observe
 for orthogonally circular polarization states in the
 fluorescence of a single trapped atom can be interpreted as
 a direct consequence of the atomic orientation or
 magnetization undergoing spontaneous or quantum
 fluctuations.

\section{Photon antibunching and transient Rabi oscillations}

 All classical fields have auto correlations $g^{(2)}(0) \ge g^{(2)}(\tau)$, and a value
 $g^{(2)}(0)-1 <0$ is classically forbidden (for classical fields one should replace all
 operators  $\hat{n}(t)$ in (\ref{eq:g2}) by their classical counterparts, intensities
 $I(t)$). This enhanced probability to detect two photons simultaneously is called 'photon
 bunching' and was observed as early as in 1956 \cite{Hanbury 56} from a usual thermal
 light source.

 For a single-atom fluorescence, however, $g^{(2)}(\tau )$ vanishes identically for $\tau
 =0$ which is a reflection of the fact that one can find at most one photon in the field
 mode of interest and can never detect two photons simultaneously. This phenomenon is
 called 'photon antibunching' and is regarded as an important manifestation of the quantum
 nature of light.

 Since the emitted light field reflects the evolution of the atomic dipole moment the
 correlation function $g^{(2)}(\tau )$ visualizes the internal dynamics of the
 observed atom for $\tau >0$. The state of an excited atom evolves continuously in the
 absence of a measurement, but theory predicts a sudden projection to the ground state
 when a photon is detected.

 This measurement 'triggers' the atom to the initial conditions
 $\rho_{gg}(0)=1$ and $\rho_{ee}(0)=0$, where $\rho_{gg}(t)$ and $\rho_{ee}(t)$ represent
 the population of the ground and excited atomic state, respectively. At that instant the
 coherent evolution starts again from the values $\rho_{gg}(0)=1$ and $\rho_{ee}(0)=0$ and
 will be interrupted by the next spontaneous emission. The normalized probability for
 detecting a second spontaneously emitted photon is now proportional to the population of
 the excited atomic state $\rho_{ee}$ according to \cite{Loudon 83}

\begin{equation}
\label{g2}
 g^{(2)}(\tau)=\rho_{ee}(\tau)/\rho_{ee}(\infty) \quad.
\end{equation}
\noindent
 Since the emission times are random, after averaging over many evolution trajectories the
 measured $g^{(2)}(\tau)$ shows directly relaxation of the system back to the equilibrium
 state after its wave function has collapsed due to the first photon detection.

\begin{figure}[hbt]
\centerline{\epsfig{file=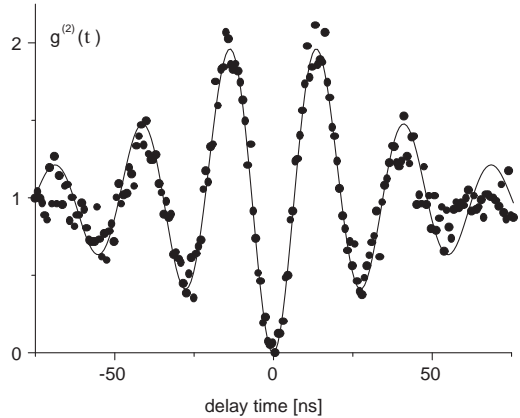,width=3in}}
    \caption{\em Intensity-intensity correlations in the resonance fluorescence of
     a single Cs atom stored in a MOT. The uncorrelated background of
     stray-light photons has been measured independently and subtracted.
     Details can be found in Ref. \cite{Gomer98a,Gomer98b}. }
    \label{fig:rabi1}
\end{figure}

 In fig. \ref{fig:rabi1} we show $g^{(2)}(\tau)$ of resonance fluorescence from a single
 atom stored in a magneto-optical trap (MOT) \cite{Raab87} with clearly observable photon
 antibunching and transient oscillations. As we already reported previously
 \cite{Gomer98a}, the observed transient oscillations in the population of the excited
 state in Fig. \ref{fig:rabi1} corresponding to coherent excitation and deexcitation
 cycles (Rabi oscillations) can be surprisingly well described by a simple model (solid
 line in Fig. \ref{fig:rabi1}) of a two-level atom (similar to $F=0 \to F=1$ transition)
 in spite of the complicated multilevel structure of the Cesium atom and light
 interference pattern (see below). This observation suggests that due to optical pumping a
 trapped atom spends most of its time in the magnetic substate that interacts most
 strongly with the local field and is forced to behave, to a good approximation, like
 a two-level system. For $\tau$ larger than the life time of the excited state the
 correlations die out due to the fluctuations of the vacuum field.

\section{Orientation dynamics of the atom revealed by polarization correlations}

 An atom trapped in a MOT and moving through the light interference pattern
 experiences various  intensities and polarizations at different places. The polarization of the resonance
 fluorescence is determined by the atomic interaction with the local light field and
 changes on the time scale of atomic transport over an optical wavelength $\lambda$. Thus,
 in addition to correlations of the total intensity one expects also polarization effects,
 that is correlations $g^{(2)}_{\alpha\beta}(\tau)$ measured between any polarization
 components $\alpha$ and $\beta$
 which should strongly depend on the atomic motion and the light-field topography.

 The light field of the MOT is formed by three mutually orthogonal pairs of
 counterpropagating laser beams with $\sigma^+$ and $\sigma^-$ polarization. A pair of two
 circularly polarized laser beams with the same handedness produces a local polarization
 that is linear everywhere with a direction of polarization that rotates a full turn every
 half wavelength. With two additional 'polarization screws' for other directions along
 with the relative phases $\phi$ and $\psi$ between these standing waves one obtains for
 the total electric field in a 3D MOT$_{\phi \psi}$

\begin{equation}
\label{eq:3DMOT}
 \vec{E}=(\sin kz + \sin ky \cdot e^{i\psi})\hat{x}+
 (\cos kz + \cos kx \cdot e^{i\phi})\hat{y}+
 (\sin kx\cdot e^{i\phi} + \cos ky \cdot e^{i\psi})\hat{z} .
\end{equation}

 We have already reported on strong correlations between circularly polarized photons and
 vanishing correlations between linear polarization components observed in the resonance
 fluorescence of a single atom trapped in a standard MOT light field configuration
 \cite{Gomer98a}. This result seems to be intuitive only for specific choices of time
 phases: for example, $\phi=90^\circ,\psi=0$ yields an 'antiferromagnetic' light-field
 structure with alternating right- and left-hand circular polarizations at points of
 deepest light shift potential, see \cite{Harald 00}. However, in a standard MOT the
 phases $\phi$ and $\psi$ change randomly due to acoustic jitter and thermal drifts. Thus
 the trapping light field has no well defined polarization state and the correlations are
 averaged over all possible values of $\phi$ and $\psi$.

 Since the MOT light field topography (\ref{eq:3DMOT}) strongly depends on the relative
 time-phases of the three contributing standing waves, we have chosen a setup
 where $\phi$ and $\psi$ are intrinsically stable \cite{Arno 97}. The
 concept uses a single standing
 wave which is multiply folded and brought into triple intersection with itself. The
 phases $\phi$ and $\psi$ can be adjusted by means of Faraday rotators. Details of this
 approach have been published elsewhere \cite{Arno 97,Harald 00}.

 If one models an atom by a "classical emitter" ($F=0 \to F=1$ transition or steady-state density matrix
 of a multi-level atom)
 then its induced dipole moment will be proportional to the local light field. This idea is often used for
 interpretation of polarization correlations in the fluorescence of a large number of laser-cooled
 atoms \cite{Jurczak 96}. As we will show this description fails completely in the case of
 a single atom.

 The most interesting case occurs for $\phi=\psi=0$. In this situation the three standing
 waves oscillate synchronously and thus the interference light field has a linear
 polarization at every point and lacks handednees completely. In this case the model of a
 classical emitter \cite{Gomer98b} predicts strong correlations between orthogonal linear
 polarization components and relatively weak correlations between circular components in
 clear contradiction with experimental results (see below).

\section{Experimental Setup}

 For polarization-sensitive correlation measurements we have trapped individual neutral
 atoms in a standard six-beam MOT \cite{Raab87} with the only exception that the phases
 and hence the light
 field topography is fully controlled \cite{Arno 97,Harald 00}. At a quadrupole field
 gradient of 12.5 G/cm the storage volume extends over approximately 100 $\mu$m. In order
 to trap small, countable numbers of atoms the loading rate from the background atomic
 vapour into the trap is kept very low. This is achieved on the one hand by lowering the
 Cesium partial pressure to $\approx 10^{-15}$ mbar (at a base pressure of $5\cdot
 10^{-10}$ mbar) and on the other hand by using trapping laser beams of diameter 4 mm
 only. The average number $\langle N \rangle $ of trapped atoms (typically between 1 and 5 in this
 experiment) can be easily adjusted by variation of the Cesium pressure. Although
 $\langle N \rangle $ also depends on the trapping laser intensity $I$ and detuning $\delta$
 of the trapping laser from atomic resonance, we are able to observe
 trapping of individual atoms over a wide range of parameters:
 $0.3 I_0 < I < 3.6 I_0$ per laser beam and $-5.2\Gamma < \delta < -1.1\Gamma$.
 The natural linewidth and the saturation intensity of
 the cooling transition are $\Gamma=2\pi\times 5.2$ MHz and $I_0=1.1$ mW/cm$^2$, respectively.

\begin{figure}[hbt]
\centerline{\epsfig{file=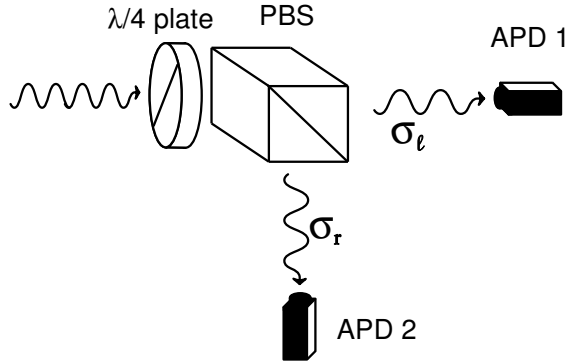,width=3in}}
    \caption{\em Optics for polarization-resolved photon correlations.
    To measure correlations between right and left handed $\ell/r$, ("circular correlations")
    a quarter wave plate is placed before the polarizing beam splitter (PBS).
    For vertical and horizontal $v/h$ ("linear correlations") components only a polarizing beam
    splitter is used. For total intensity-intensity correlations the fluorescence is divided by a
    non-polarizing beam splitter. }
    \label{fig:poloptics}
\end{figure}

 The atomic resonance fluorescence is due to excitation by the trapping laser field only.
 Fluorescent light is collected from a $5\%$ solid angle by a lens and then splitted into
 orthogonal polarization states by means of polarizing optics (see Fig.
 \ref{fig:poloptics}) which also directs the corresponding light onto two avalanche photo
 diodes (APD). The APDs are operated in single photon counting mode and achieve a photon
 detection efficiency of 47 \% at a  dark count rate of 10 s$^{-1}$. The average photon
 count rate for an individual atom lies in the range 3-10 kHz in our experiments,
 depending on laser intensity and detuning. Observation direction is in the xy-plane at
 45$^\circ$ to the laser beams (the $z$-axis is the symmetry axis of the MOT quadrupole
 magnetic field).

 Usually, a measurement of the cross correlation function along with the total
 intensity-intensity correlation provides complete information: the corresponding auto
 correlation can be inferred from the sum rules for orthogonal polarization components
 \cite{Gomer98b}. For example for circular components one has
 $g^{(2)}_{++}(\tau)+g^{(2)}_{+-}(\tau)=2g^{(2)}(\tau)$. A cross correlation measurement
 ($+-$) can be carried out 4 times faster than the corresponding auto correlation
 measurement ($++ $) where only one half of the total fluorescence is detected.

 A computer registers the arrival times of all photons from the two APD channels with 100
 ns time resolution and with 700 ns dead time in each channel. The entire experimental
 information accessible in the set-up is stored and thus can be processed afterwards by
 correlation analysis through numerical multi-stop procedures, which completely eliminate
 systematic errors such as photon pile-up introduced by single-stop methods traditionally
 used in Hanbury Brown \& Twiss type experiments \cite{Hanbury 56,Coates68}.

 Atoms are randomly loaded from background vapour and randomly lost due to collisions with
 background gas. But
 since individual atom arrival and departure events are easily located within 1 ms, it is
 straightforward to determine the instantaneous number of atoms from the average count
 rate. Note that the number of trapped atoms fluctuates on the second time scale
 \cite{Gomer98b}. This enables us to separate all data from a single experimental run into
 different classes with the number of trapped atoms as a parameter - the data for
 different atom numbers are therefore obtained under identical experimental conditions. It
 is also easily possible to distinguish correlations of the fluorescence of trapped
 individual atoms from uncorrelated background of detection events due to stray light or
 fluorescence from thermal, untrapped atoms. As a consequence all measured correlation
 functions can always be unequivocal normalized.

\section{Experimental Results}

\begin{figure}[hbt]
\centering \epsfig{file=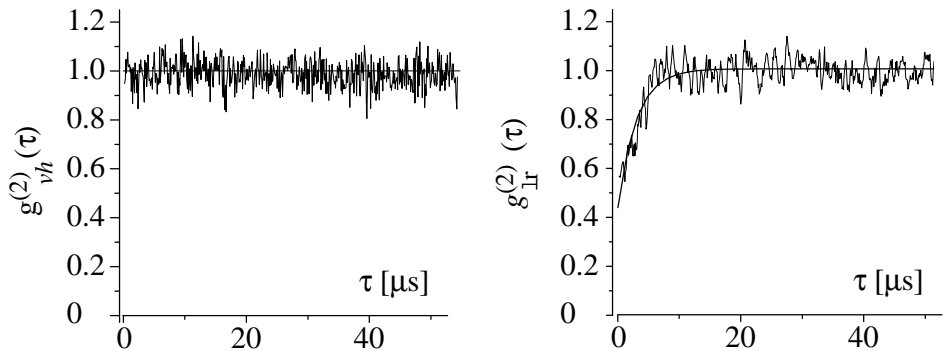, width=12.5cm}
 \caption{\em  Measured photon correlations from a single atom in the MOT$_{00}$ field.
               Left: Correlation function $g^{(2)}_{vh}(\tau)$ for orthogonal linear
               polarization components ($\delta=-1.8 \Gamma$, $I=0.95 I_0$, integration time 6.3 min).
               Right: Correlation function $g^{(2)}_{\ell r}(\tau)$ for orthogonal
               circular polarization components ($\delta=-2.7 \Gamma$, $I=0.7 I_0$, integration time 11.1 min).
               Solid line: exponential fit.}
 \label{fig:pol_corr}
\end{figure}

 Measured  second order correlation functions for orthogonal polarization components
 in the fluorescence of a single atom in the MOT$_{00}$ are shown in  fig. \ref{fig:pol_corr}.
 Within our experimental uncertainties correlations are completely absent for
 linear polarization components of the fluorescence, in baffling contrast with the result
 for circularly polarized components.

\begin{figure}[hbt]
\centerline{\epsfig{file=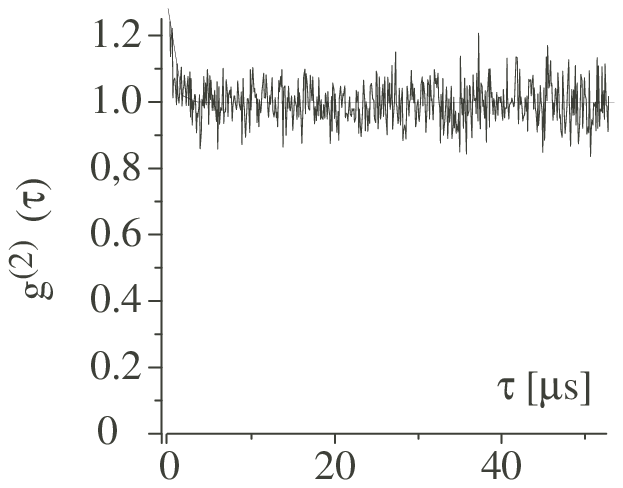}}
    \caption{\em Total intensity-intensity correlation function $g^{(2)}(\tau)$ of a single-atom fluorescence
                 in the MOT$_{00}$ field ($\delta=-1.1 \Gamma$, $I=1.3 I_0$, integration time 10.4 min).
                 Solid line: exponential fit.}
    \label{fig:int_corr}
\end{figure}

 For a single atom the circular correlation contrast reaches
 s values up to 62\%. This very strong correlation typically
 relaxes within a few $\mu$s. Since photon antibunching and
 Rabi oscillations at ns time scales are not resolved in
 these measurements, we have assumed for a simple analysis
 the relaxation process of the polarization correlations to
 be exponential and extracted a single relaxation time
 constant $\tau_r$. The intensity-intensity auto
 correlation $g^{(2)}(\tau)$ also shows a contrast of about 30\% due to intensity
 modulations of the MOT light field, but the relaxation time
 constant of about 0.6 $\mu$s in fig. \ref{fig:int_corr} is
 significantly shorter than for the circular cross
 correlations \cite{Raithel 98}.

\begin{figure}[hbt]
\centerline{\epsfig{file=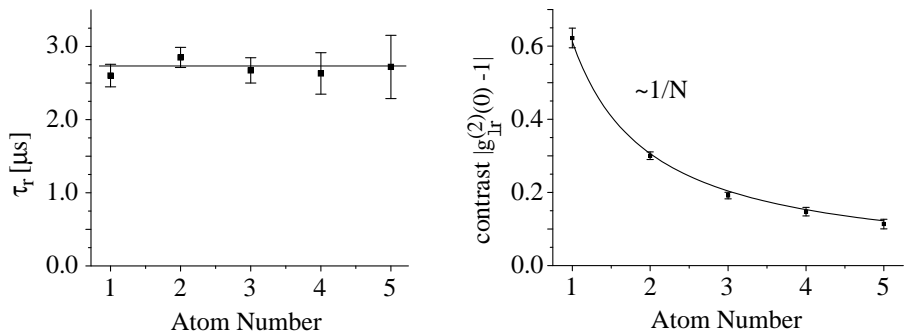, width=12cm}}
    \caption{\em Measured characteristic relaxation time (left) and contrast (right) of the cross
    correlations for circular polarization components as a
    function of the number of observed atoms recorded at constant trap parameters.}
    \label{fig:ndepend}
\end{figure}

 Furthermore, we have experimentally verified that $\tau_r$
 does not depend on the number of trapped atoms  and that
 the contrast of the correlation function is proportional to
 the inverse number of atoms, $N^{-1}$ (see fig.
 \ref{fig:ndepend}). Thus, as one would expect, we deal with
 a pure single-atom effect.

\section{Discussion}

\begin{figure}[hbt]
\centerline{\epsfig{file=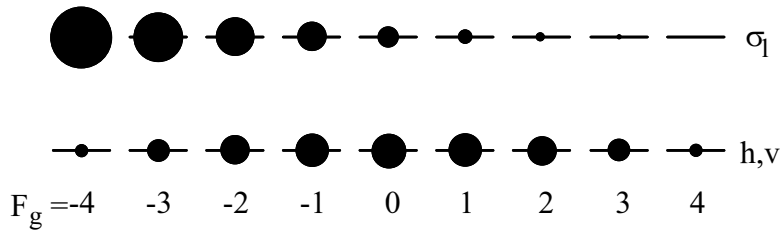}}
\caption{\em
 Ground state population after decay of an initially isotropic exited state distribution with
 $\langle m \rangle =0$ by emitting a circular polarized photon $\sigma_l$ (upper part).
 The dot sizes correspond to the individual populations of the sublevels.
 For comparison, the situation is completely different for the $hv$-correlation: Detecting
 a linearly h- or v-polarized photon (lower part) does not lead to any orientation of the atom.
 Therefore any succeeding  excitation will result in almost equal
 probabilities for detecting a $h$- or a $v$-polarized photon in the next spontaneous
 emission, leading to $g^{(2)}_{hv}(\tau=0) \approx 1$.
 }
\label{fig:pops}
\end{figure}

 In order to interpret the polarization properties of the resonance fluorescence of atoms
 driven by a light field with linear polarization we begin by considering an atom at rest.
 As this light field does not favor any of the two circular polarization states, one can
 assume that prior to the detection of the first photon the distribution of atomic
 magnetic sublevels is symmetric, $\langle m \rangle = 0$. If we suppose for simplicity
 that the local light field consists of equal parts of both orthogonal circular
 polarizations only, it is clear that the $\langle m \rangle = 0$ state is unstable
 \cite{Forston 96}. At the level of an individual particle this equilibrium state can be
 distorted by the observation of a single circularly polarized photon, which projects the
 atom into its ground state, breaks the symmetry of the Zeeman substate population
 (fig. \ref{fig:pops}), and
 creates an imbalance in the interaction strengths with both circular polarization
 components. The next absorption will be preferentially further enhance the asymmetry. The
 imbalance in the interaction strengths rapidly grows with $m$ leading to fast pumping
 into one of the outmost Zeeman states $m=\pm F$. The ratio of the interaction strengths
 in these stretched  states reaches the value $(2F+1)(F+1)$ making them very stable for
 large $F$ even in the presence of the polarization component in the light field driving
 $\Delta m=0$ transitions. The atom in this oriented state prefers to radiate into the
 same polarization state as the first detected photon, resulting in anticoincidences in
 the cross correlation for orthogonal circular polarizations.

 Note that this effect is a specific feature of $F \to F+1$ transitions and does not occur
 for $F \to F$ or $F \to F-1$ transitions. It can be regarded as bistability, since an
 emission of a circularly polarized photon makes the atom more likely to be pumped in the
 corresponding outmost Zeeman state. As both stretched states $m = + F$ and $m = - F$ are
 equivalent in the presence of a linear polarized light field, this effect is principally
 unobservable in an atomic ensemble. However, if one adds some circular polarization
 component to the driving light field a spontaneous spin polarization of a macroscopic sample can be
 observed as recently demonstrated in \cite{Forston 96} by optically pumping on the $F=3
 \to F=4$ hyperfine component of the $D_1$-line of Cs.

 Although the real 3D-situation in our experiment is much more complicated than the simple
 model presented above, our interpretation is furthermore supported by the following
 considerations: As noted in \cite{Forston 96}, the condition for magnetic bistability in
 a linearly polarized light field is given by $|90^\circ - \beta|< 45 ^\circ$, where
 $\beta$ is the angle between the magnetic field (quantization axis) and the light
 polarization. In the MOT$_{00}$ there are 8 intensity antinodes in a unit cell with
 directions of the local linear polarization at these points coinciding with the diagonals
 of the coordinate system $(\pm1,\pm1,\pm1)$. It is easy to see that the bistability condition is
 fulfilled for the majority of points in the MOT magnetic quadrupole field $B\propto
 (-x,-y,2z)$.

 However, in our case the stretched states are not intrinsically stable. While radiation pressure
 forces are balanced for an aligned atom with $\langle m \rangle$=0, they are unbalanced for
 an oriented atom since the local linearly polarized light field is created by (at least two)
 {\em counterpropagating} laser beams with orthogonal circular polarizations.
 The imbalance in the light forces created by atomic orientation thus causes acceleration,
 or heating which is again damped by the usual laser friction forces \cite{Dalibard89}.
 Thus for our experiment we must acknowledge that the observation of a circularly polarized photon
 not only redefines atomic orientation but also its mechanical status: Internal and external
 atomic degrees of freedom are inextricably entangled.

 The measured relaxation time constant of $g^{(2)}_{\ell r}(\tau)$ indeed depends strongly
 on the atom-light field interaction which also governs atomic motion in the trap. The
 interaction strength is measured by the light shift parameter $\Lambda$  \cite{Coh 92,Harald 00}
 corresponding to the maximum energy shift of the atomic energy levels.  Under our
 experimental parameters sub-Doppler cooling leads to atomic temperatures proportional to
 the light shift, thus $T_{kin} \propto \Lambda$ \cite{Raab87,Dalibard89,Ste 91}.

\begin{figure}[hbt]
\centerline{\epsfig{file=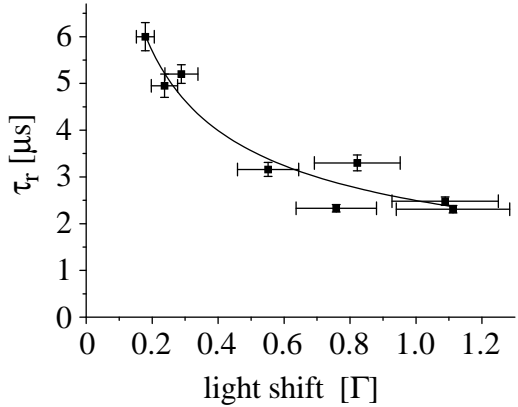}}
    \caption{\em The relaxation time constant $\tau_r$ as a function of the
             light shift parameter $\Lambda$. Solid line: fit function $\propto \Lambda
             ^\alpha$, with $\alpha=-0.51\pm0.01$.}\label{fig:tauvslsh}
\end{figure}

 In fig.\ref{fig:tauvslsh} we show the measured dependence of the time constant $\tau_r$
 as a function of $\Lambda$. As expected we find good agreement of our experimental data
 with the functional relationship $\tau_r\propto 1/\bar{v} \propto 1/\sqrt{T_{kin}} \propto
 1/\sqrt{\Lambda}$.  Here ${\bar{v}}$ denotes the average atomic velocity. This means that
 relaxation of the spontaneous atomic magnetization is determined by atomic motion through
 the light field.

 We can carry out our analysis one step further if we assume that the characteristic length
 over which relaxation takes place is $\tau_r\overline{v}$ = $\lambda/2$, the spatial period
 of the MOT light field. It is then straightforward to evaluate
 characteristic kinetic temperatures from our correlation measurements to be in the range
 between 10-68 $\mu$K for the $\Lambda$ range shown in fig. \ref{fig:tauvslsh}, in good
 agreement with previous measurements \cite{Raab87,Ste 91}.

\section{Summary}

 Photon correlations observed in the resonance fluorescence of a single atom provide a
 direct access to the internal atomic dynamics. We have shown two examples of resolved
 quantum dynamics of an isolated atom stored in a magneto-optical trap. Beyond observation
 of the well-known phenomenon of transient Rabi oscillations (usually connected with
 photon antibunching), we have observed fluctuations of the atomic magnetic orientation by
 measuring photon correlations between orthogonal polarization components.

 Using a simple model we have given evidence for the following dynamical processes causing
 strong circular cross correlations in resonance fluorescence: Spontaneous emission of
 circularly polarized photons causes instantaneous orientation. Subsequent photons are
 preferentially absorbed and emitted with identical polarization. This memory effect
 leading to correlated absorption of photons with equal polarization and thus to increased
 momentum diffusion of an atom with a multi-level structure
 has been discussed in \cite{Dalibard89,Castin 90}. In our experiments
 we can clearly isolate this effect by observation of an anticorrelation of
 circularly polarized photons successively emitted with opposite handedness.

 Subsequent optical pumping induced by atomic motion in the light field causes relaxation of
 the orientation clearly seen in the photon correlations. In a sense we have seen the
 elementary sub-Doppler cooling and heating forces in a
 $\sigma^+\sigma^-$-molasses at work.

 We thank Svenja Knappe for providing some basic techniques in an early stage of the
 experiment. We are also pleased to acknowledge Frans E. van Dorsselaer and  Gerhard
 Nienhuis for fruitful discussions and sharing their insight into elementary atomic
 processes. This work is supported by the Deutsche Forschungsgemeinschaft (DFG).

\end{document}